\documentclass[useAMS,usenatbib]{mn2e}
\usepackage {graphicx}
\onecolumn
\title[The role of resonances in barred galaxies]{The role of resonances in barred galaxies}
\author[D. Ceverino and A. Klypin]{D. Ceverino$^{1}$ and A. Klypin$^{1}$\\
$^{1}$New Mexico State University}
\begin{document}


\maketitle


\begin{abstract}

The dynamics of the inner disk of many galaxies is dominated by bars. As a result, resonances
between the bar and the disk become an important factor. 
In order to detect resonances, we measure angular and radial frequencies of individual orbits in N-body models of barred galaxies.
 In our simulations with several million particles,  stellar disks are immersed in live halos of dark matter. A bar forms 
spontaneously and interacts with the disk through resonances. 
We detect 9 of them. The corotation (CR) and the inner Lindblad resonance (ILR) are the most significant.
 The spatial distributions of particles trapped at these resonances show that ILR is not
 localized at a given radius. Particles trapped at ILR lie along the bar spanning a wide range of radii. 
On the other hand, particles at the corotation resonance form a ring around the corotation radius.
 We study how orbits evolve near corotation and how they are trapped. 
When an orbit is getting trapped, it evolves strongly, but after that there is little evolution. 
Those trapped orbits circulate along the corotation ring or librate around a stable lagragian point.   
The same technique is applied  to study resonances in the halo. 
The bar interacts strongly with the halo particles that stay close to the disk.
We find  the corotation resonance and  the inner Lindblad  resonance for halo particles.

\end{abstract}

This paper contains the figures presented in the conference: "Dynamics of Galaxies: Baryons and Dark Matter'' (Las Vegas, NV, March 2005).
The full talk is available in http://astronomy.nmsu.edu/dept/html/directory.grads.danielcv.shtml
\begin{figure*}
{\includegraphics[width =0.47\textwidth]  {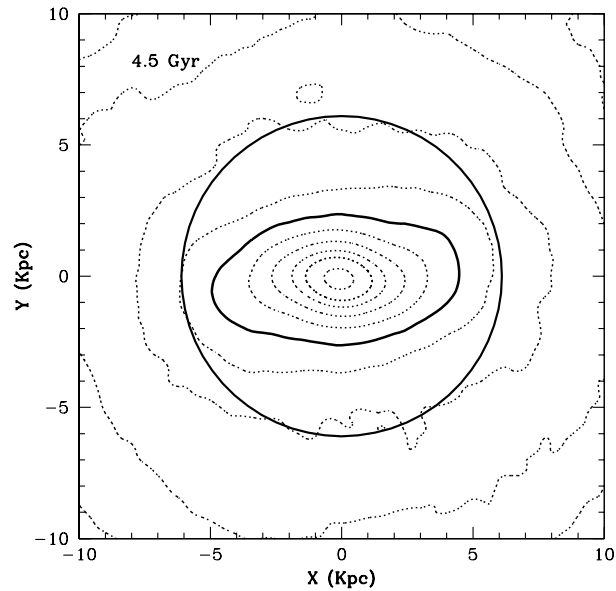}}
 \caption{Face-on view of a N-body model of a barred galaxy with a strong bar at 4.5 Gyrs (MODEL 1). 
The plot shows contours of equal surface density. 
The circle represents the corotation radius for this model. }
\end{figure*}
\twocolumn
\begin{figure}
{\includegraphics[width =0.47\textwidth]{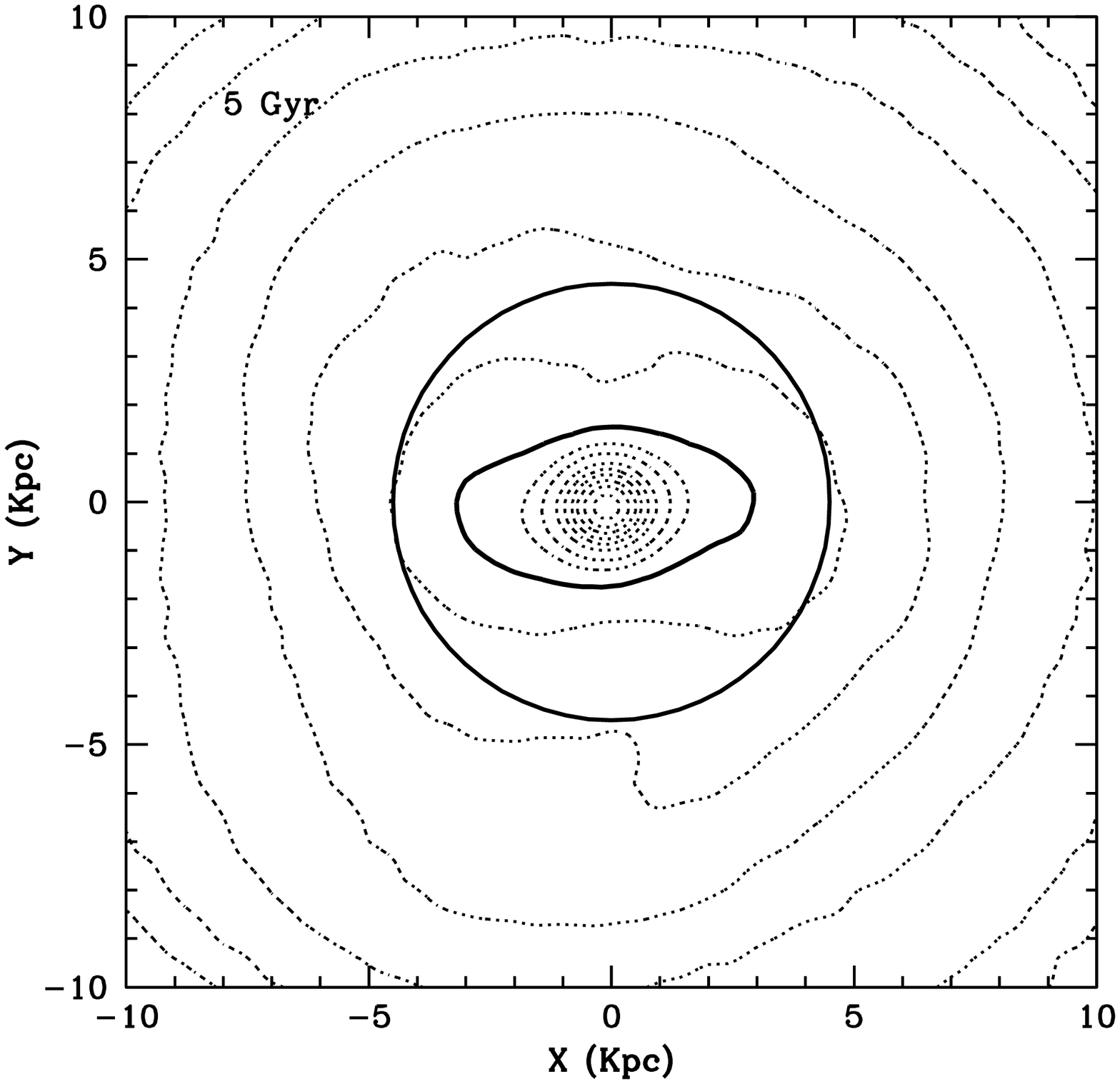}}
 \caption{MODEL 2: Face-on view of a N-body model of a barred galaxy with a weak bar and a highly concentrated center (pseudobulge)
The plot shows contours of equal surface density. 
The circle represents the corotation radius for this model. }
{\includegraphics[width =0.47\textwidth]{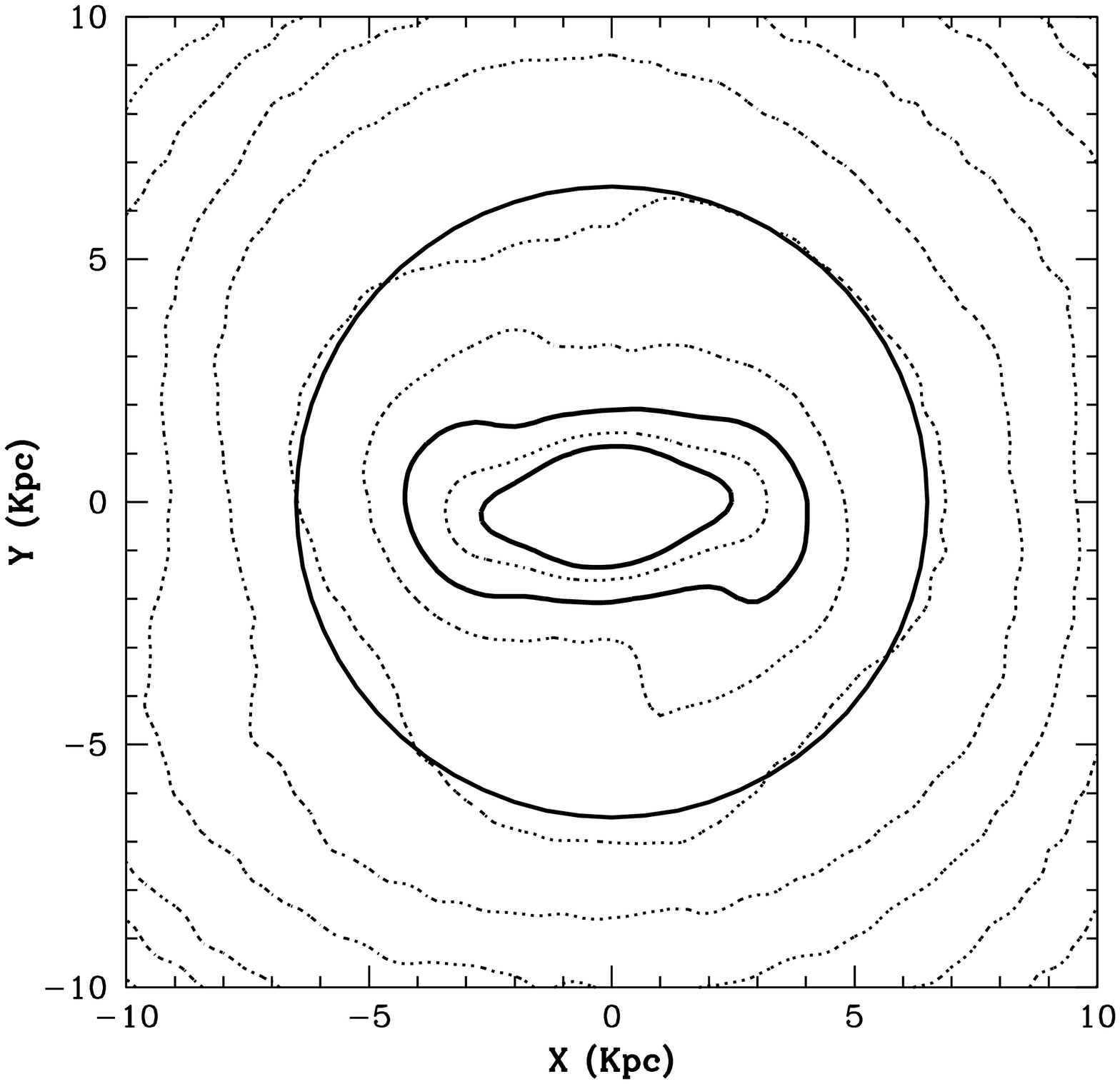}}
  \caption{MODEL 3: Model C of Valenzuela and Klypin 2003. The plot shows contours of equal surface density. 
The circle represents the corotation radius for this model. }
\end{figure}  
\begin{figure}
{\includegraphics[width =0.47\textwidth]{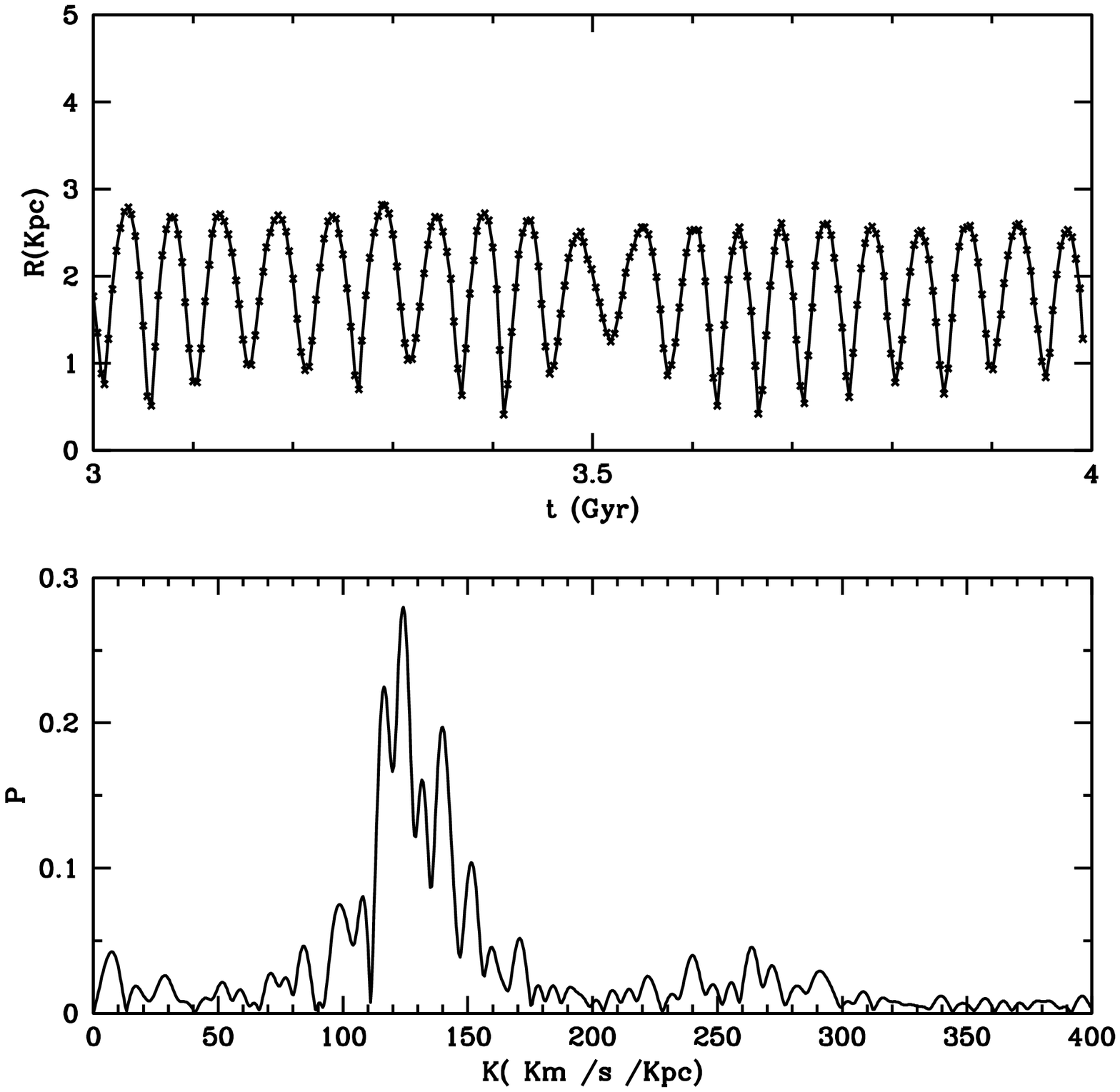}}
 \caption{Example of a radial frequency $(\kappa)$  measurement for a particle using the radial evolution of its trajectory and its power spectrum. 
The top panel shows the radial evolution of a particle in model 1 and the bottom panel presents the Fourier decomposition of its radial evolution.
The radial frequency is measured as the maximum peak of the spectrum $ ( \kappa =126 Km / s / Kpc )$. 
}
{\includegraphics[width =0.47\textwidth]{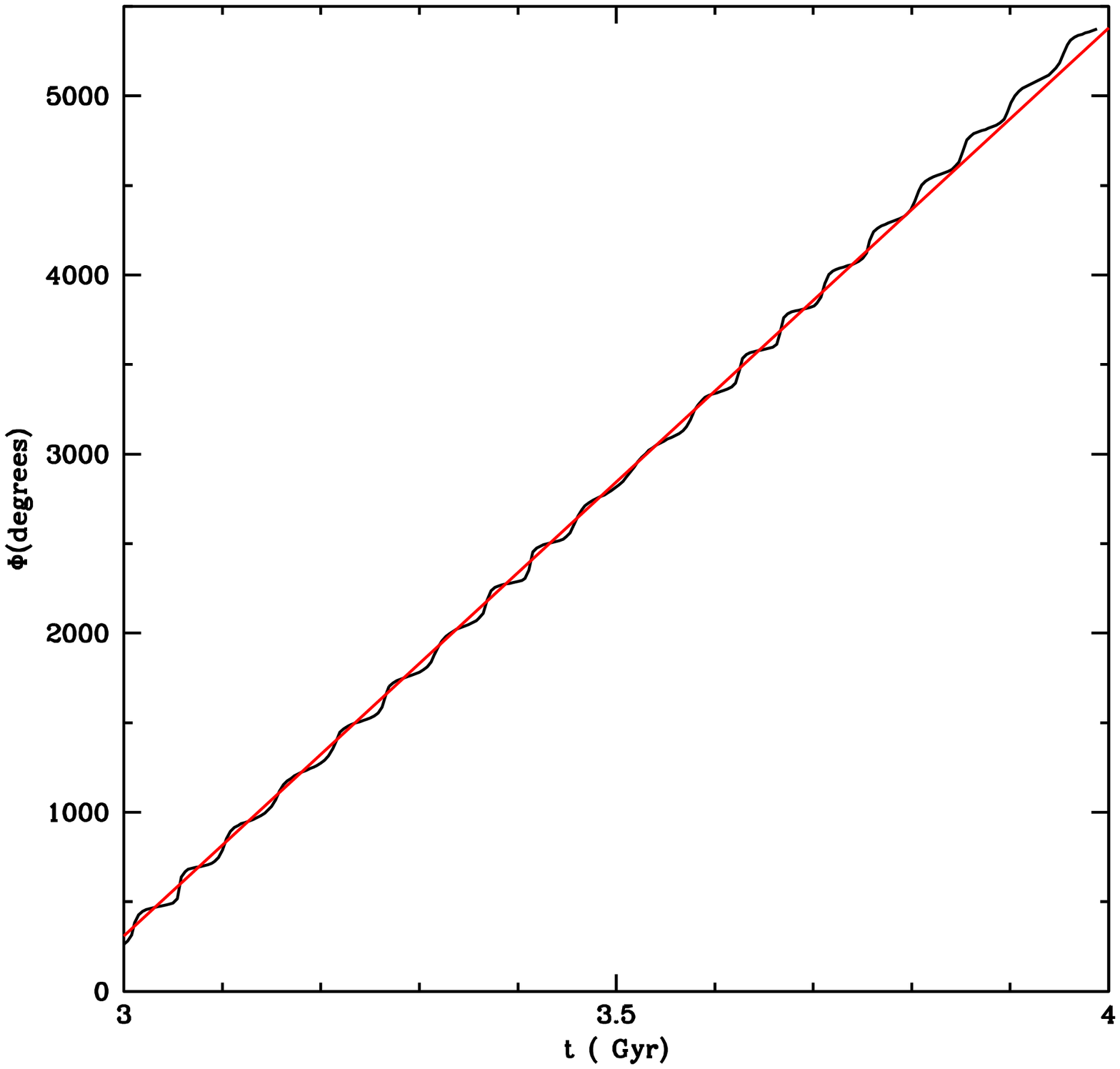}}
 \caption{Example of a angular frequency $(\Omega)$ measurement for a particle  using its angular position along its trajectory. 
The figure presents the angular evolution of the particle selected in model 1 with $\Omega = 88 Km / s / Kpc $. 
The straight line shows the slope of this angular frequency.    }
\end{figure}  
\begin{figure}
{\includegraphics[width =0.47\textwidth]{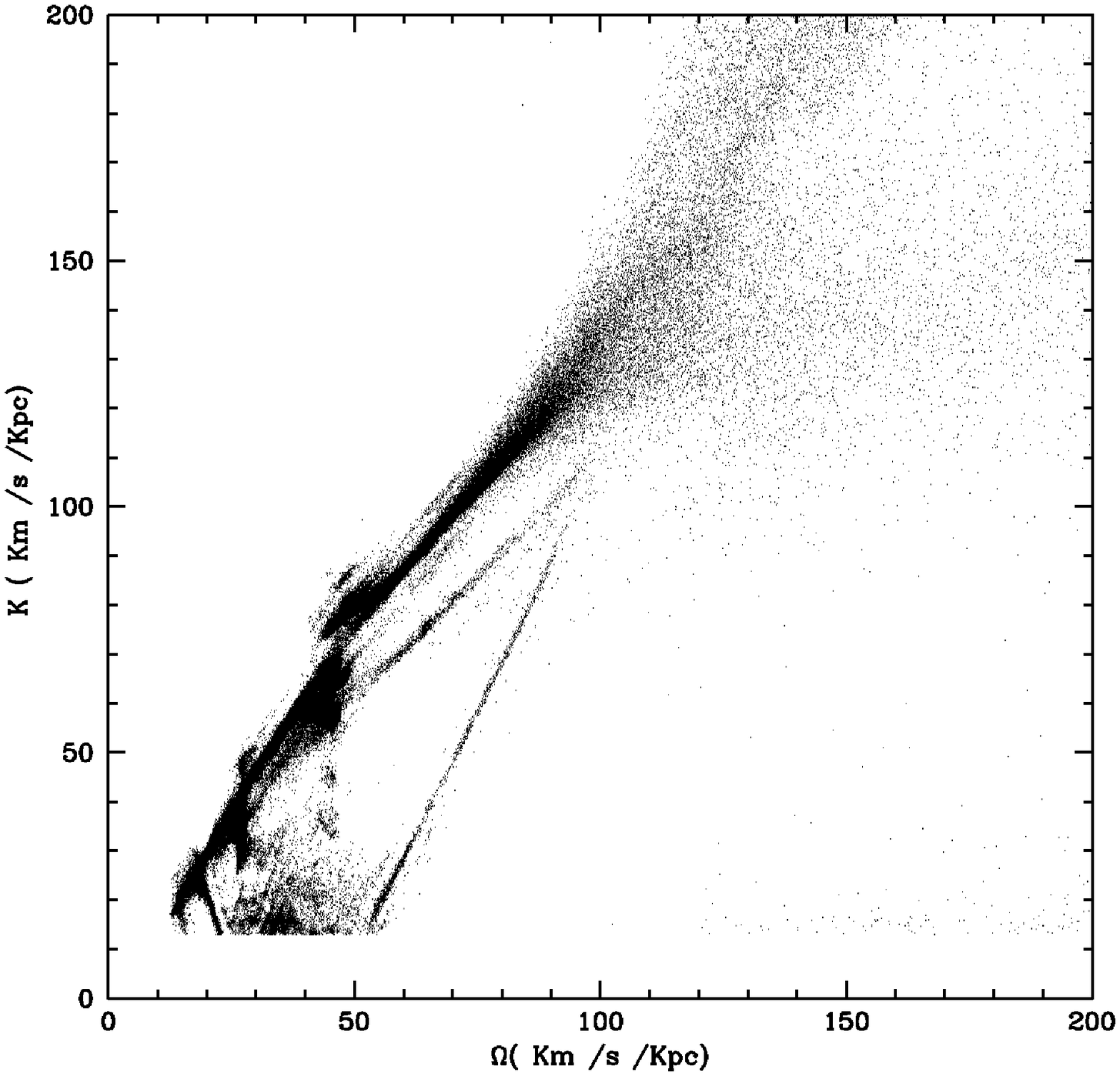}}
 \caption{Distribution of particles in the frequency space for model 2. Horizontal axis shows angular frequencies $(\Omega)$ and vertical axis shows radial frequencies $(\kappa)$. Each point represents an orbit of a particle during 1 Gyr of evolution. 
 Clustering of particles along straight lines with certain slopes indicates the presence of resonances. }
{\includegraphics[width =0.47\textwidth]{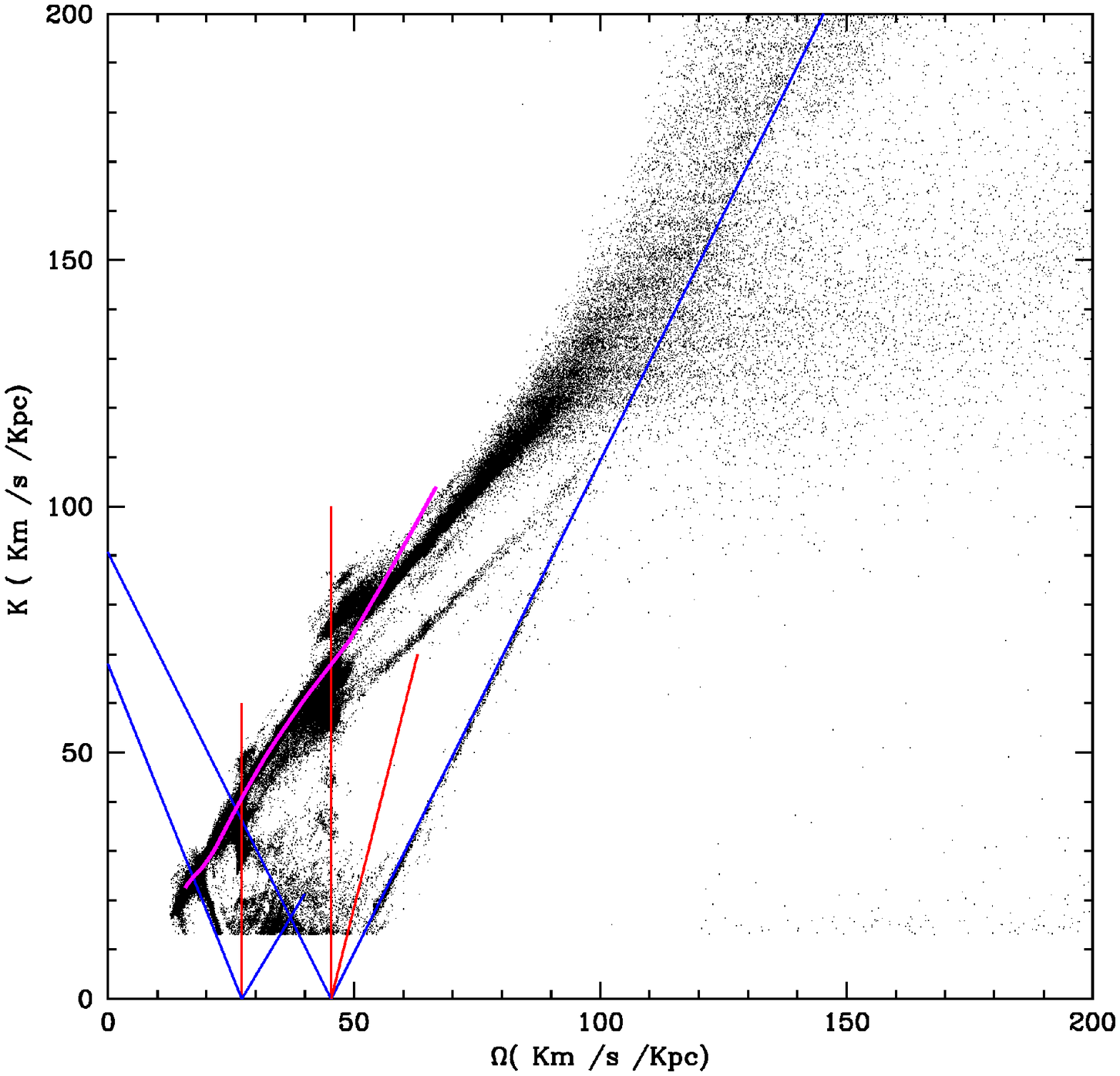}}
 \caption{ The same plot showing different resonances as straight lines. The inner Linblad resonance (ILR) is the rightmost line and 
the corotation resonance (CR) is the rightmost vertical line.
 The magenta curve is the result of the epicycle approximation which breaks down inside corotation, where the orbits are very elongated.}
 \end{figure}  
\begin{figure}
{\includegraphics[width =0.47\textwidth]{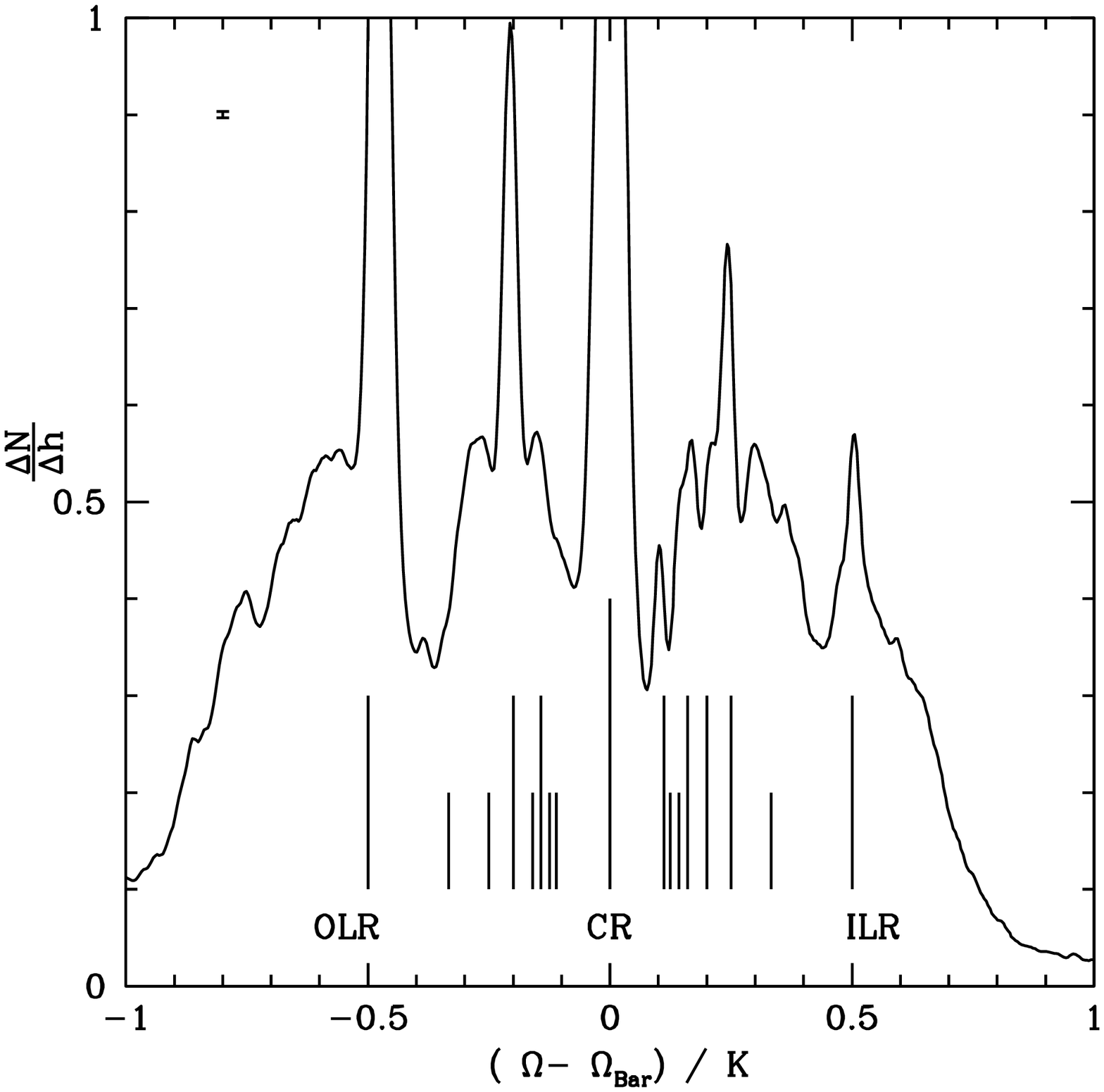}}
 \caption{Distribution of the ratio $ ( \Omega - \Omega_{B} ) /  \kappa$ for the model 2 for 1 Gyr of evolution. 
The vertical axis shows the fraction of particles per unit bin.
 Vertical lines represent resonances from -1:2 (ILR) to -1:9 (right side) and from 1:2 (OLR) to 1:9 (left side). CR lies at the center of the plot. 
The peaks show a strong indication of trapping resonances and no indication of gaps at low order resonances.
 The errorbar at the upper-left corner is the $1\sigma$ error using poison noise.}
{\includegraphics[width =0.47\textwidth]{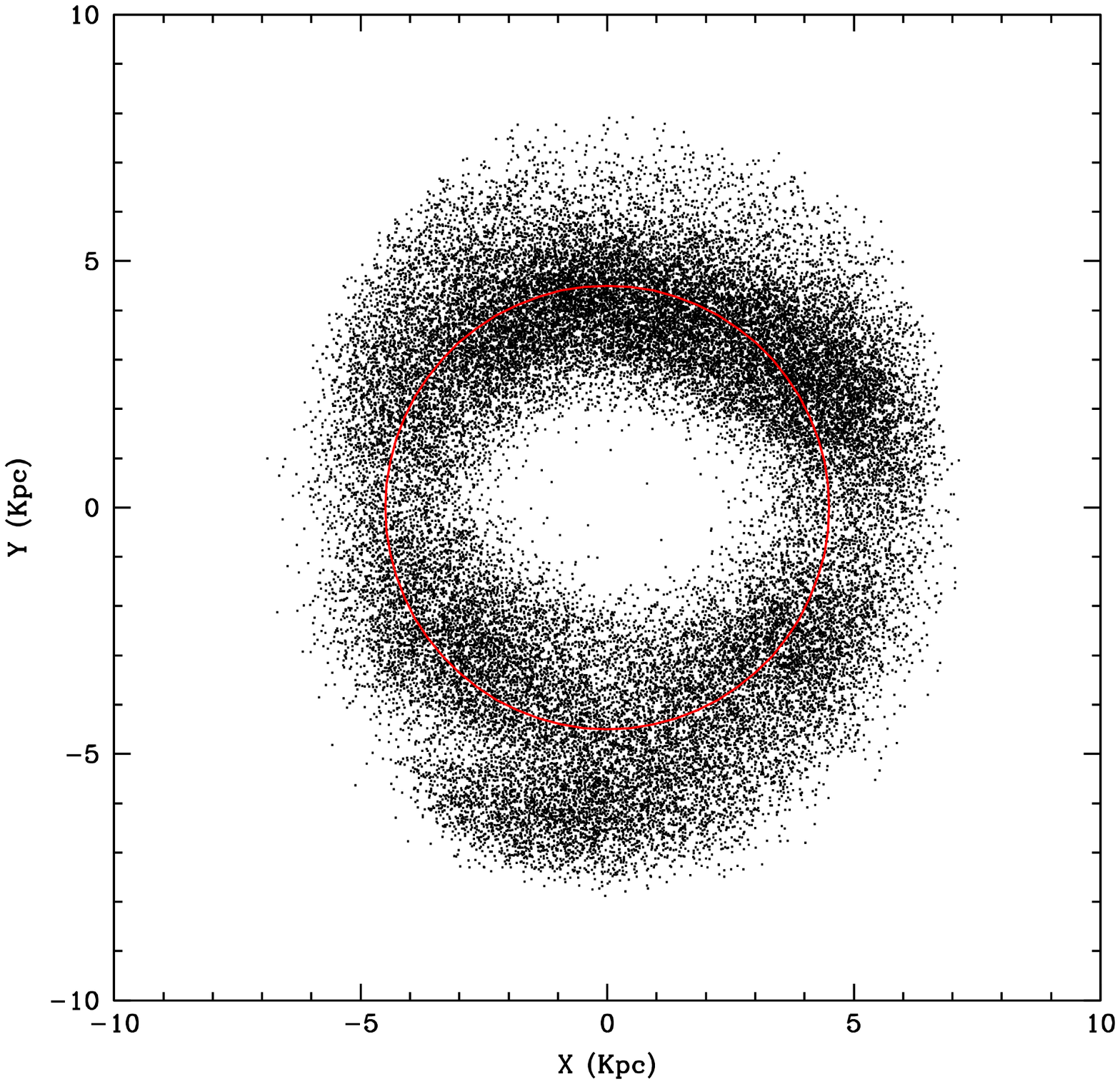}}
 \caption{Spatial distribution of particles at the corotation resonance in the model 2. These particles lie along the corotation ring represented by a circle. The bar stays parallel to the horizontal axis.  }
\end{figure}  
\begin{figure}
{\includegraphics[width =0.47\textwidth]{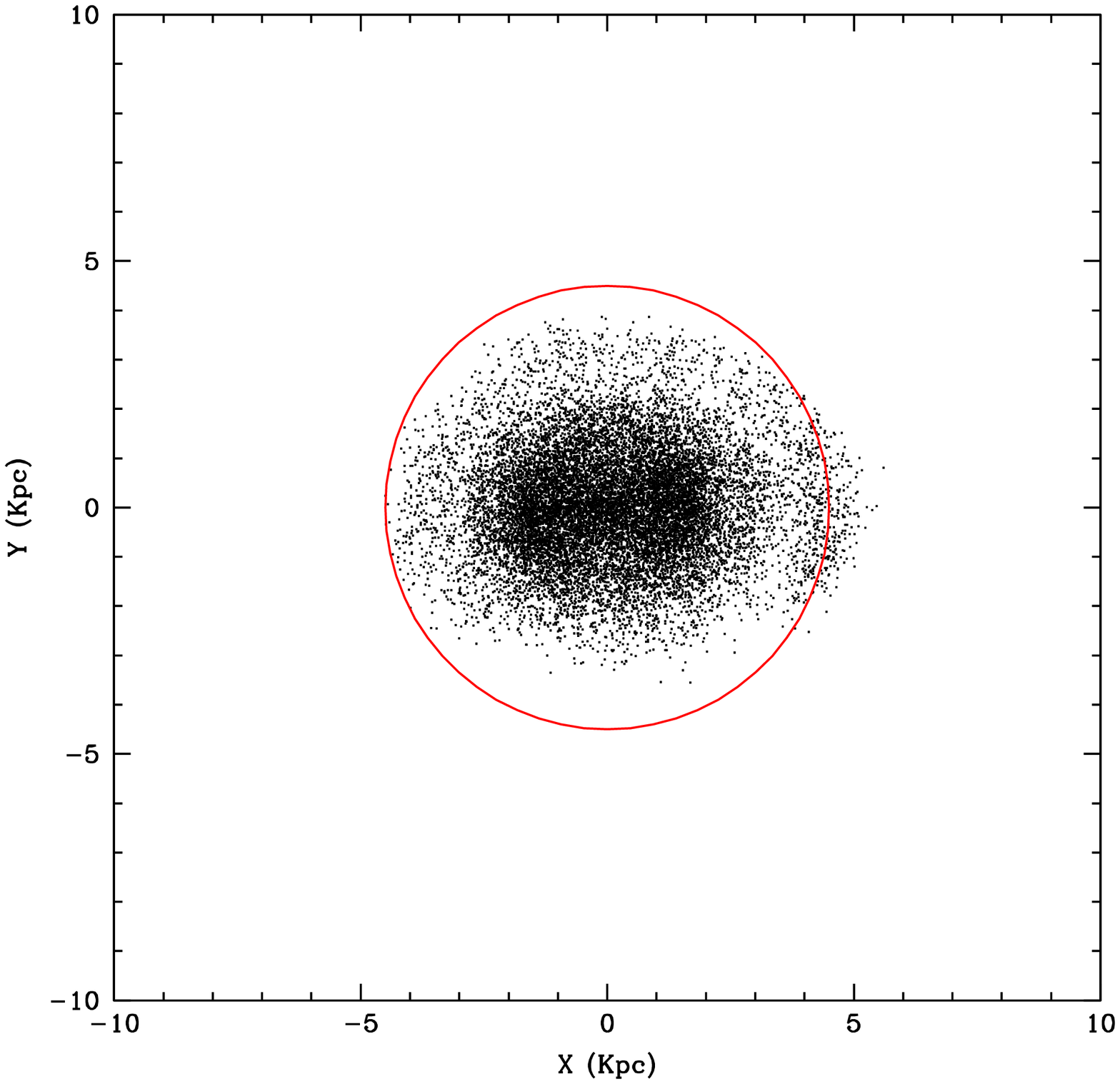}}
 \caption{Spatial distribution of particles at the inner Linblad resonance in the model 2. 
The distribution is not localized at any radius, but it is spread along the bar. 
ILR particles can come arbitrary close to the center.
The circle shows the corotation radius. The bar stays parallel to the horizontal axis.  }
{\includegraphics[width =0.47\textwidth]{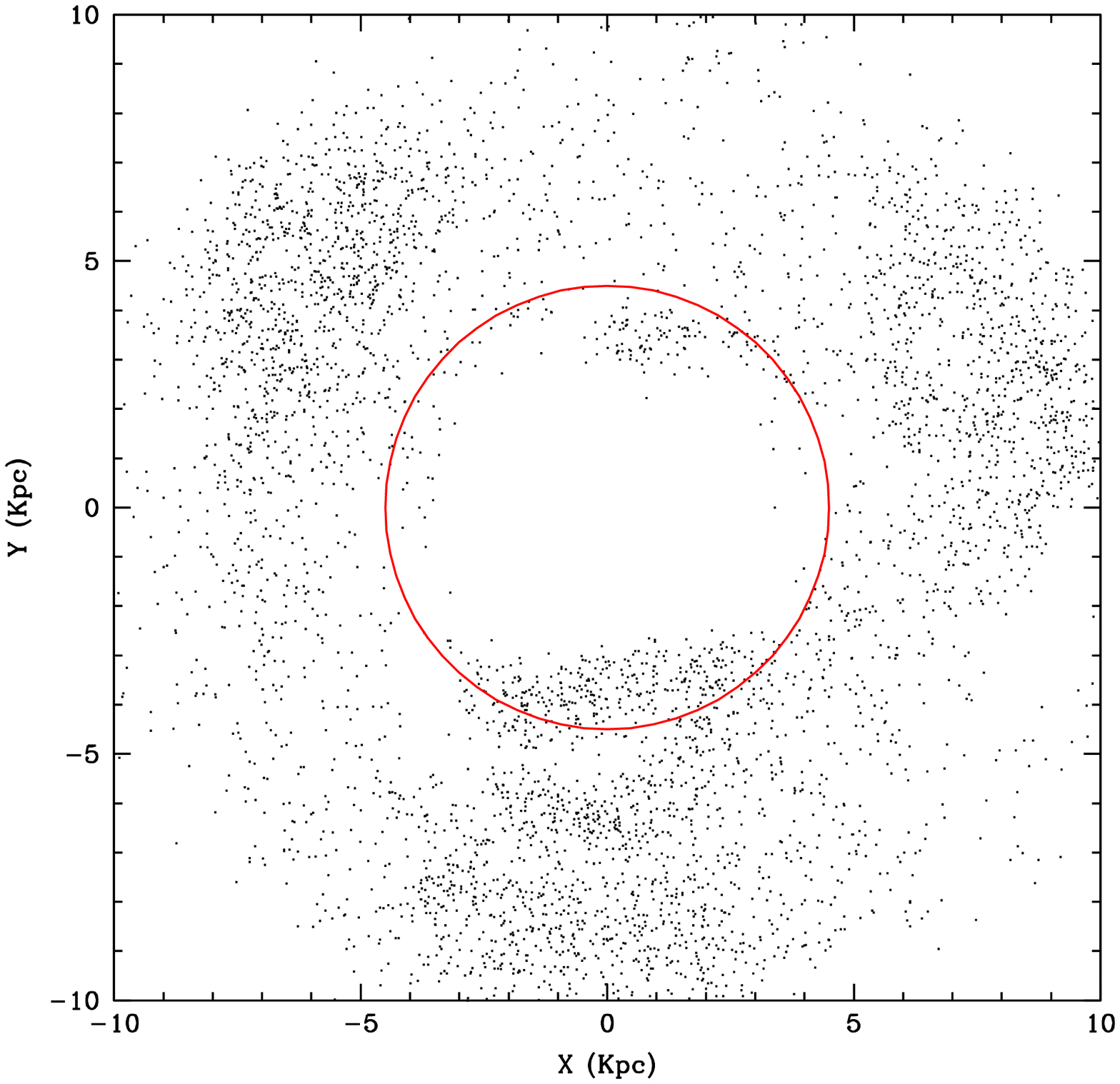}}
 \caption{Spatial distribution of particles at the outer Linblad resonance in the model 2. The circle presents the corotation radius.  }
\end{figure}  
\begin{figure}
{\includegraphics[width =0.47\textwidth]{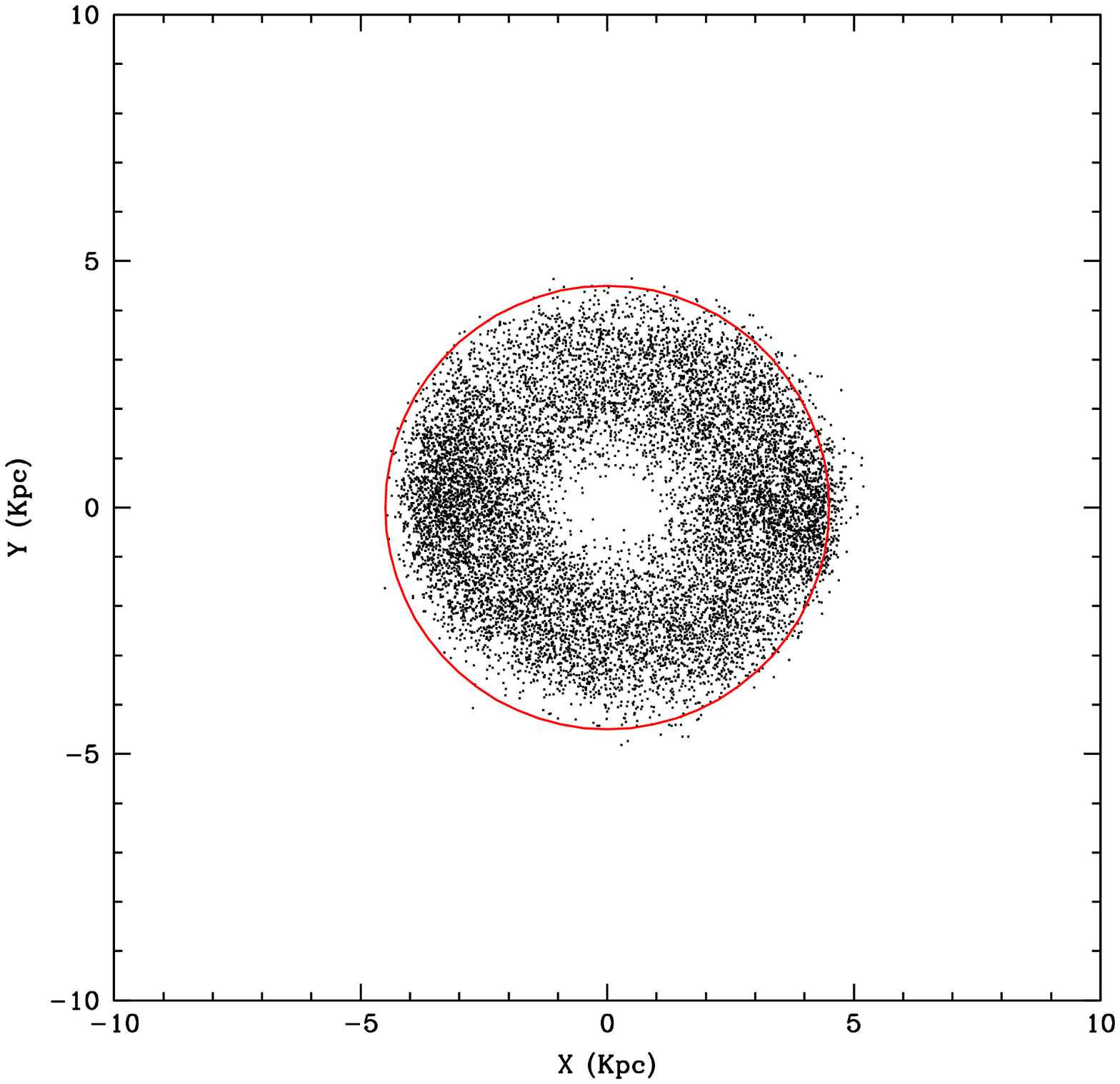}}
 \caption{ Spatial distribution of particles at the ultra-harmonic resonance (-1:4) in the model 2. The circle presents the corotation radius.  }
{\includegraphics[width =0.47\textwidth]{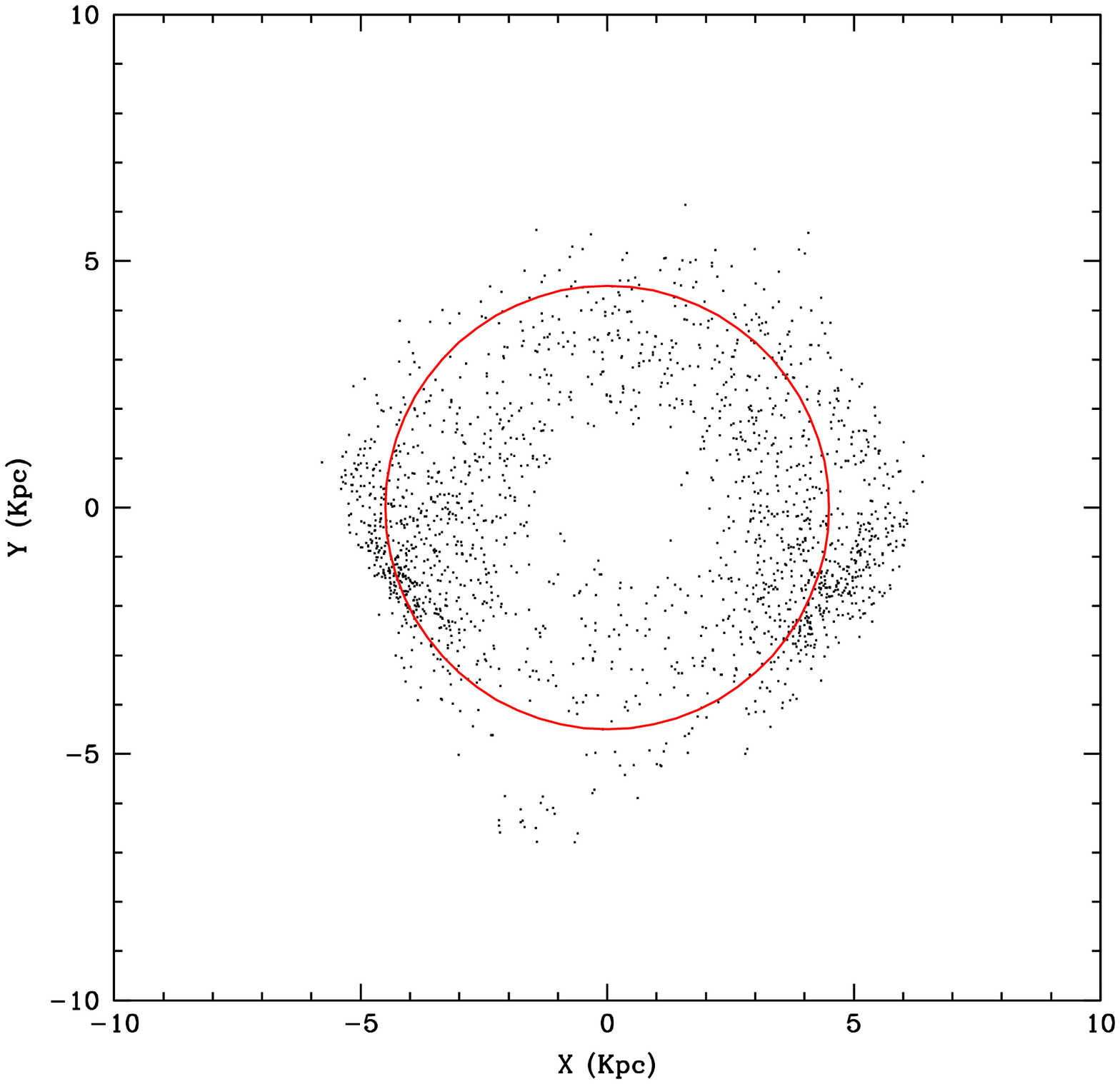}}
 \caption{Spatial distribution of particles at the -1:9 resonance in the model 2. The circle presents the corotation radius. }
\end{figure}  
\begin{figure}
{\includegraphics[width =0.47\textwidth]{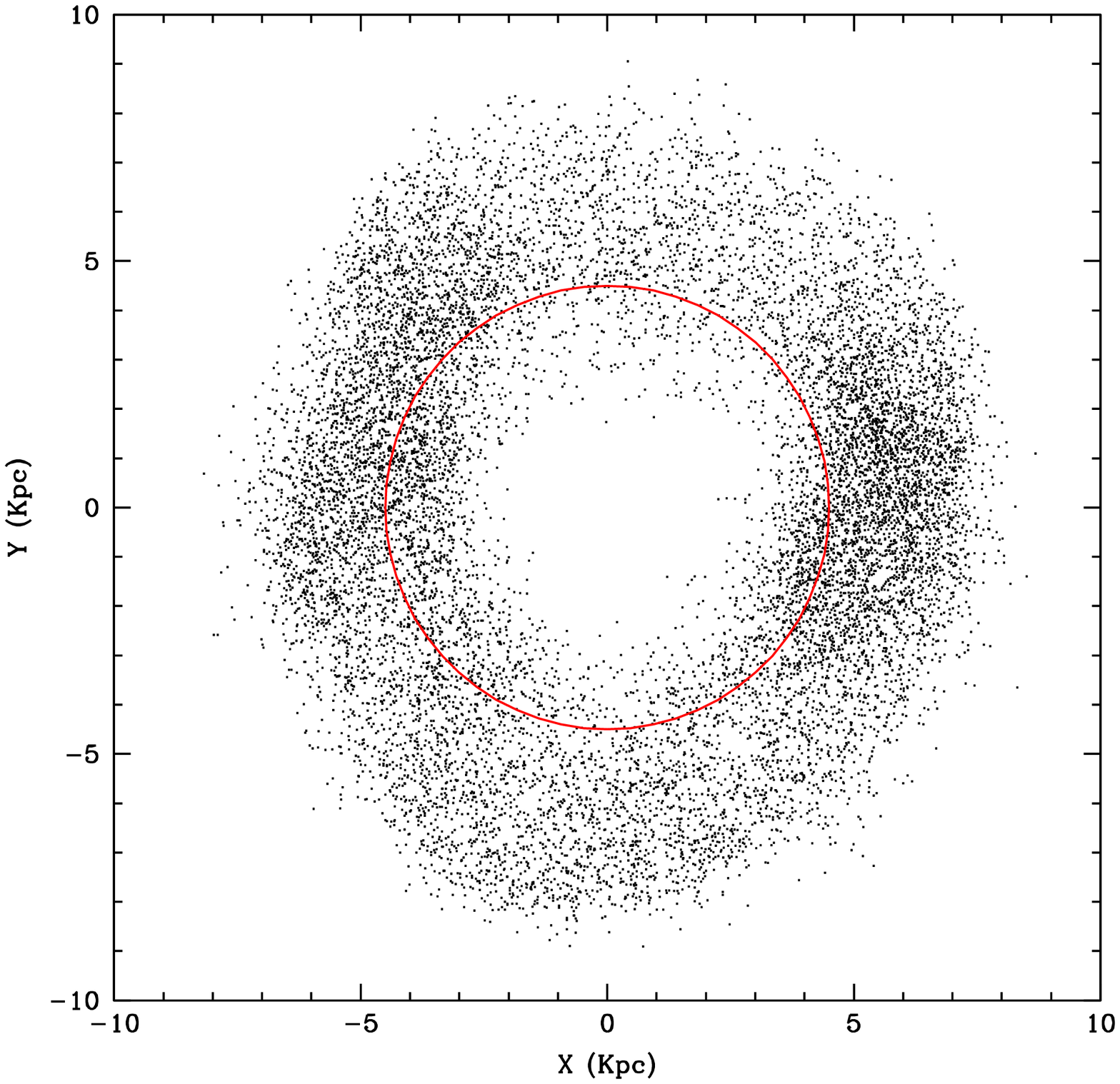}}
 \caption{Spatial distribution of particles at the 1:5 resonance in the model 2. The circle presents the corotation radius.  }
{\includegraphics[width =0.47\textwidth]{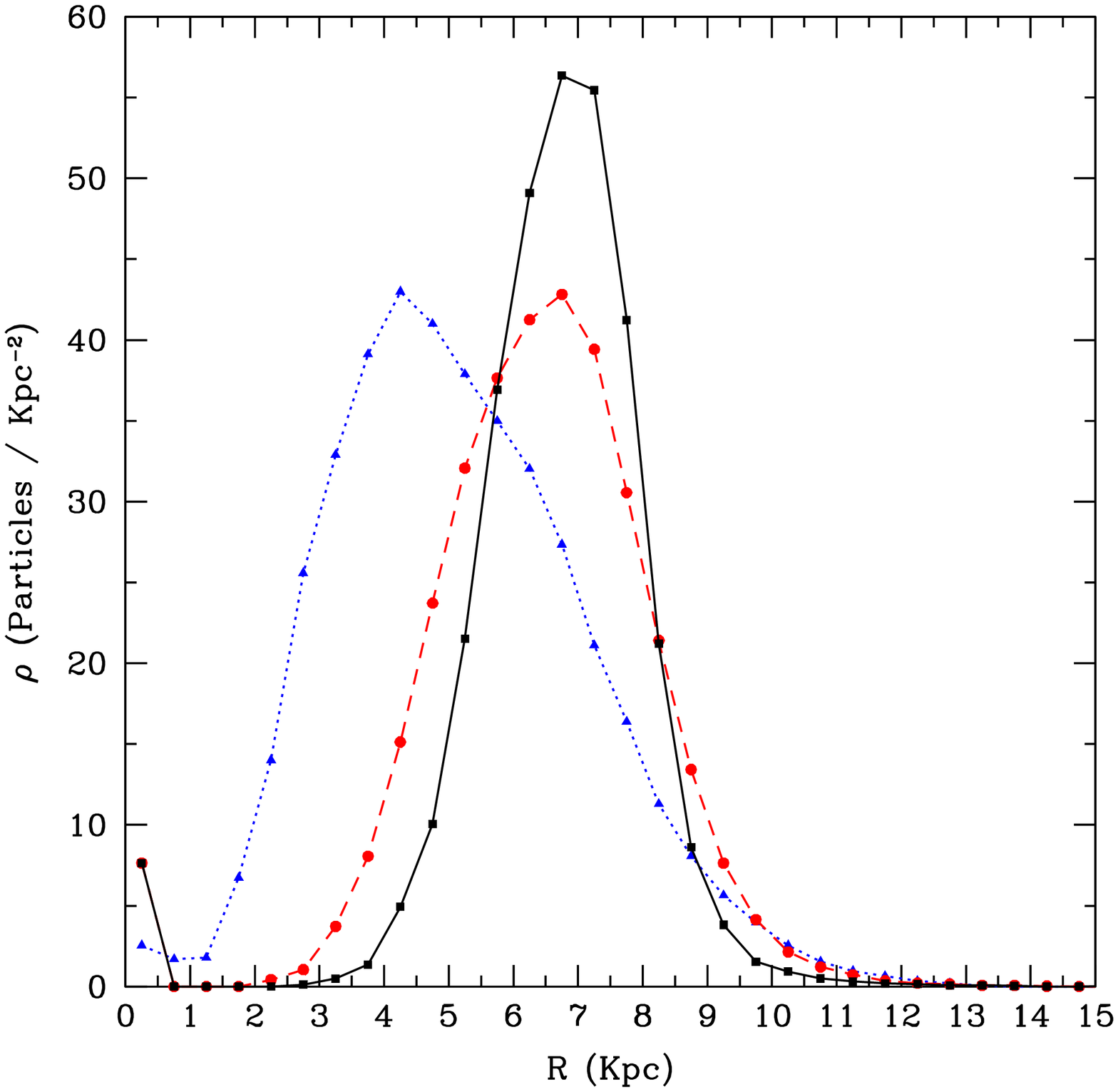}}
 \caption{ Evolution of the surface density profile of particles which were at the corotation resonance at 3.5 Gyr for model 3.
 The full curve correspond to the distribution of the particles at 3.5 Gyr. 
 The dot curve shows the same particles at 0.1 Gyr ( before the bar formation). The dash line is for 4.5 Gyr. 
As the bar forms, some particles get trapped near corotation radius, forming a stable ring.}
\end{figure}  
\begin{figure}
{\includegraphics[width =0.47\textwidth]{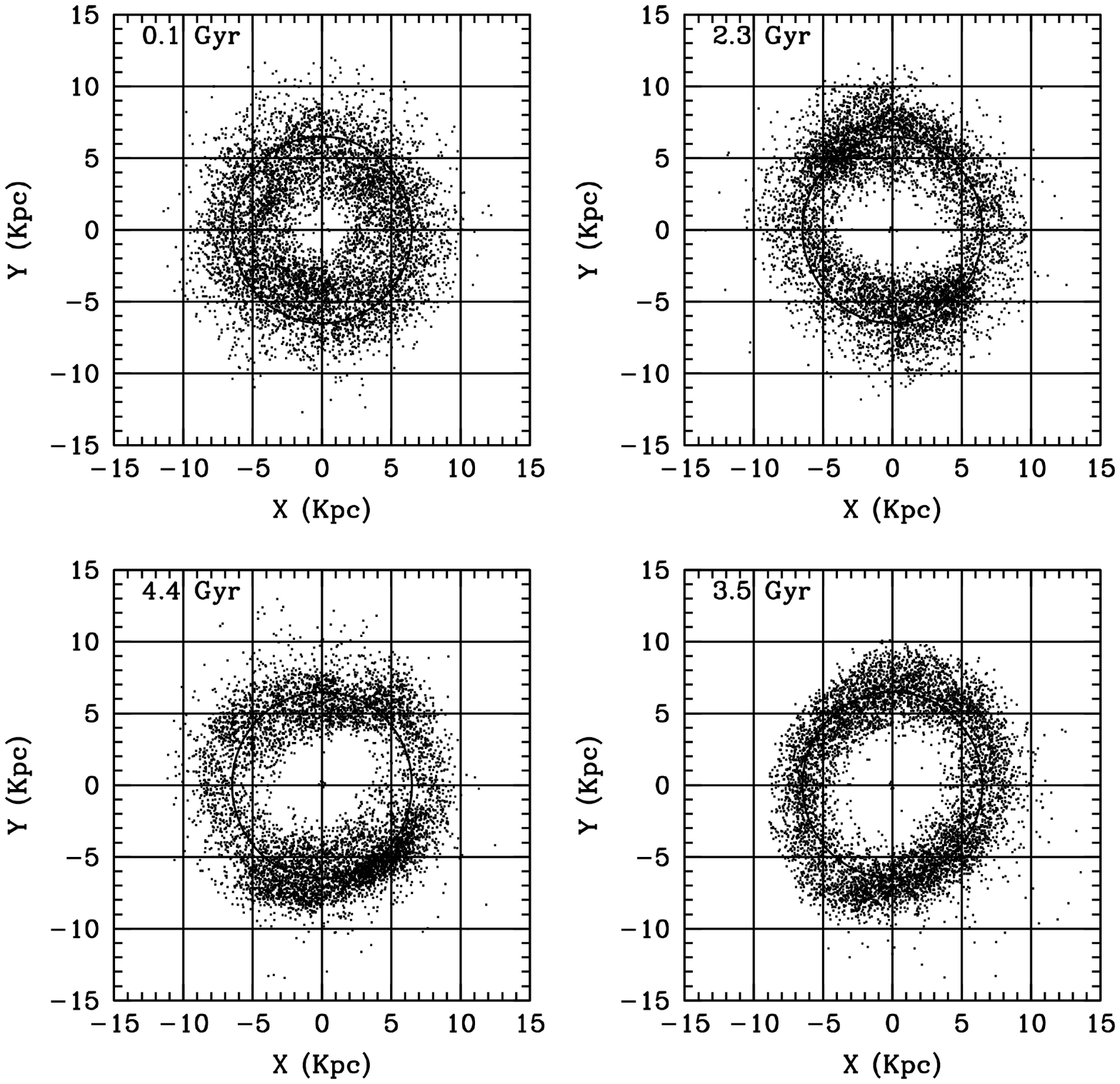}}
 \caption{ Evolution of the spatial distribution of  particles, which were at the corotation resonance at 3.5 Gyr for model 3. The circle represent the corotation radius. Particles trapped near corotation stay there for a long time. }
 {\includegraphics[width =0.47\textwidth]{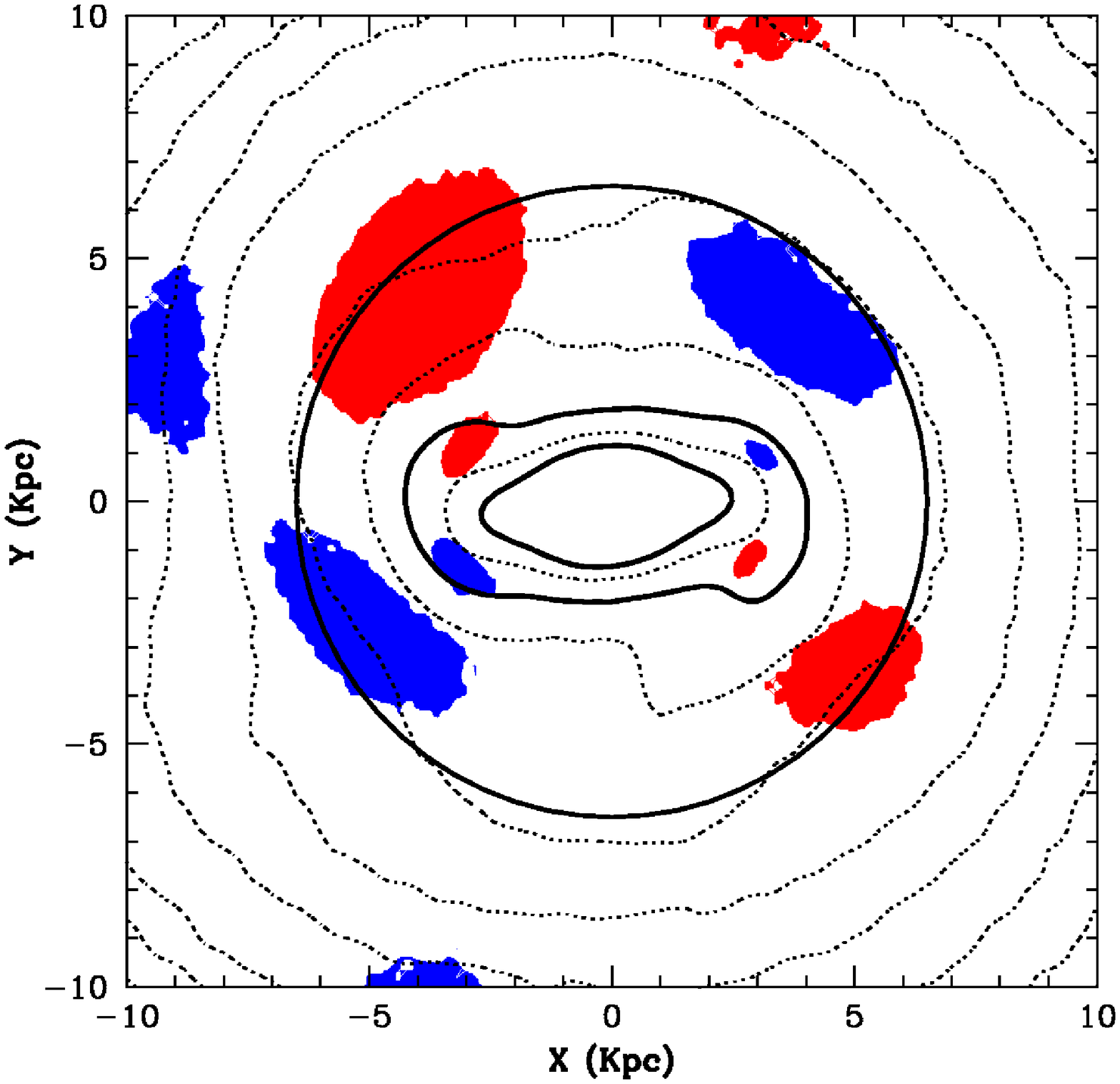}}
 \caption{ Distribution of the change in angular momentum in the disk for a short period of time (0.15 Gyrs). 
The gray areas show the position of the particles with the maximum  increment in their angular momentum.
The black areas show the position of the particles with the maximum decrement of angular momentum.
 Those areas lie mainly along the corotation radius (black circle)}
\end{figure}  
\begin{figure}
 {\includegraphics[width =0.47\textwidth]{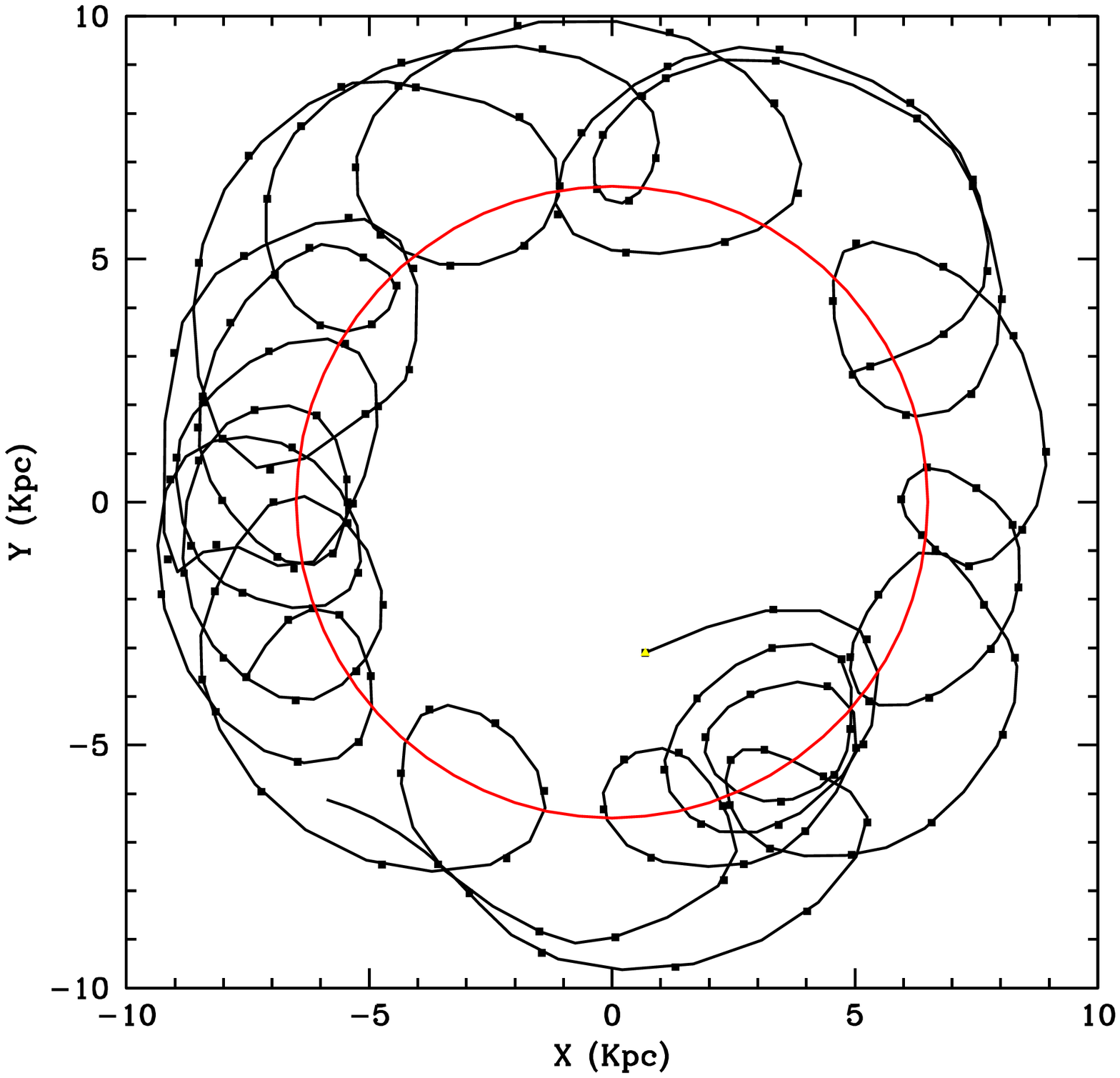}}
 \caption{Example of a trajectory near the corotation resonance. The particle circulates around the corotation radius (circle) after it got trapped. 
 The reference frame rotates with the pattern speed.  So, the bar stays always along the horizontal axis. }
{\includegraphics[width =0.47\textwidth]{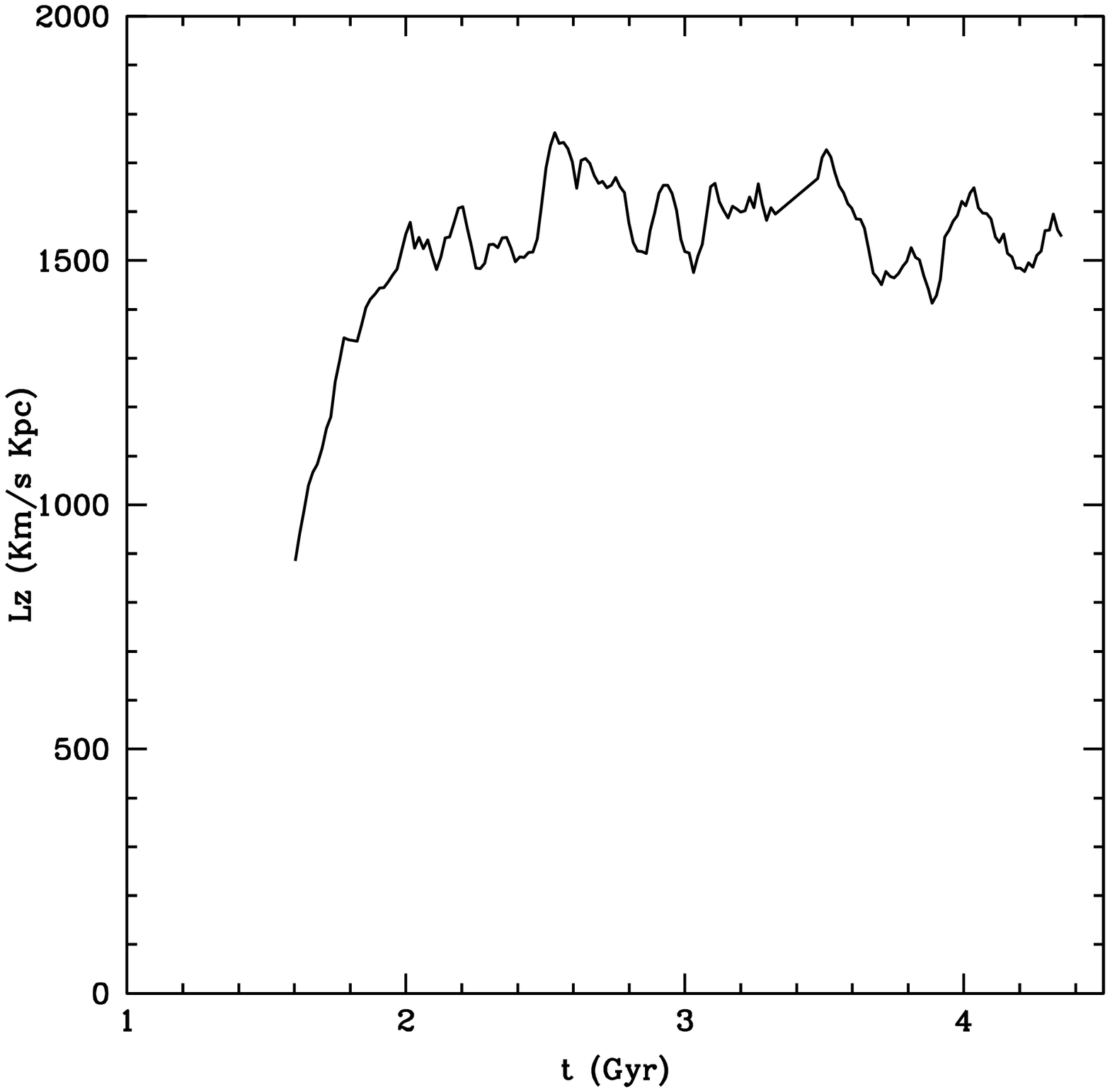}}
 \caption{ Angular momentum history of the same particle shown in figure 18. The particle increases its angular momentum when it gets trapped at the corotation resonance, but after that the angular momentum does not change significantly although the particle stays close to the corotation resonance.}
\end{figure}  
\begin{figure}
 {\includegraphics[width =0.47\textwidth]{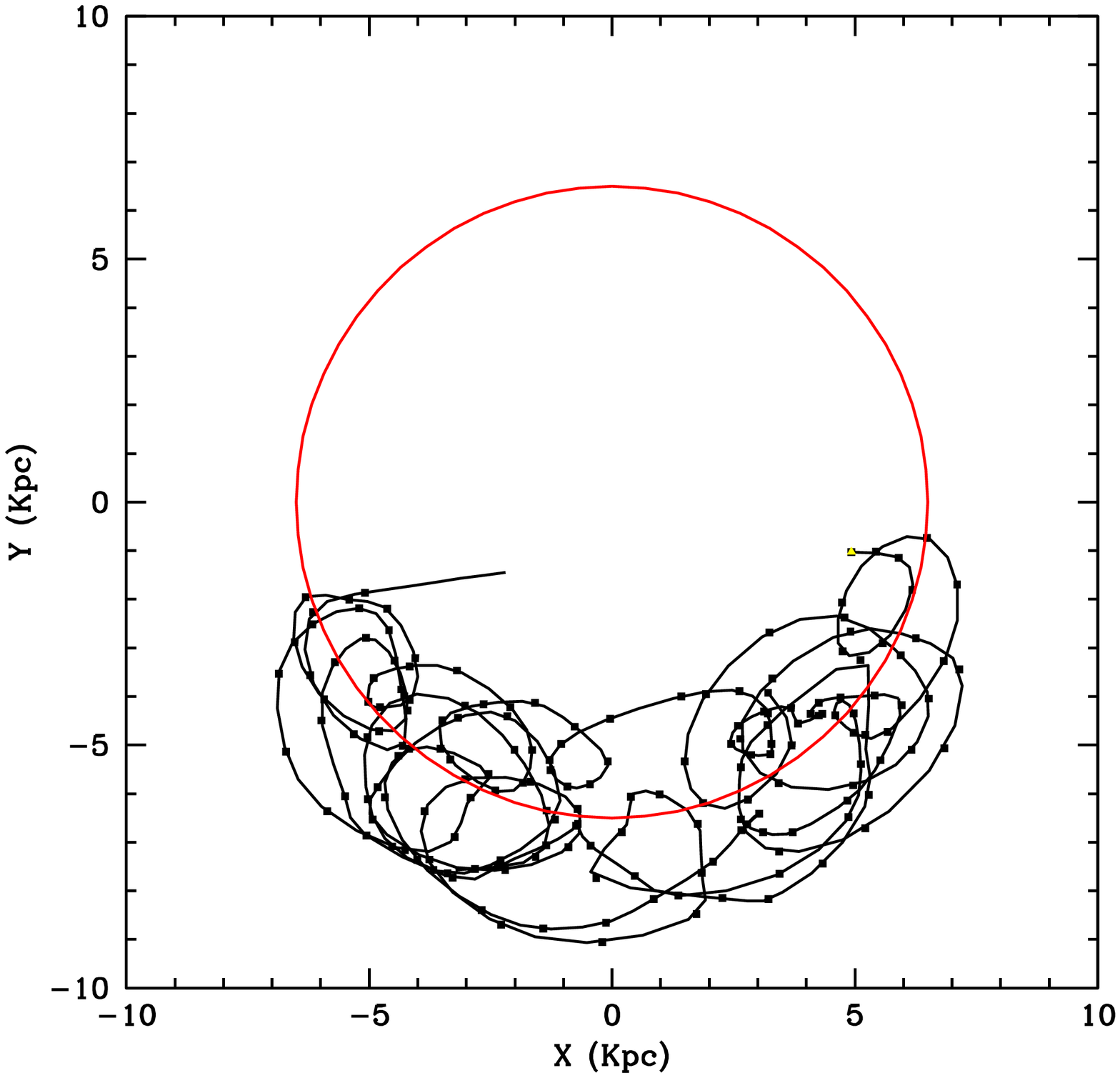}}
 \caption{ Example of libration around the corotation radius. The particle librates around the areas of maximum positive and negative change of the angular momentum.}
{\includegraphics[width =0.47\textwidth]{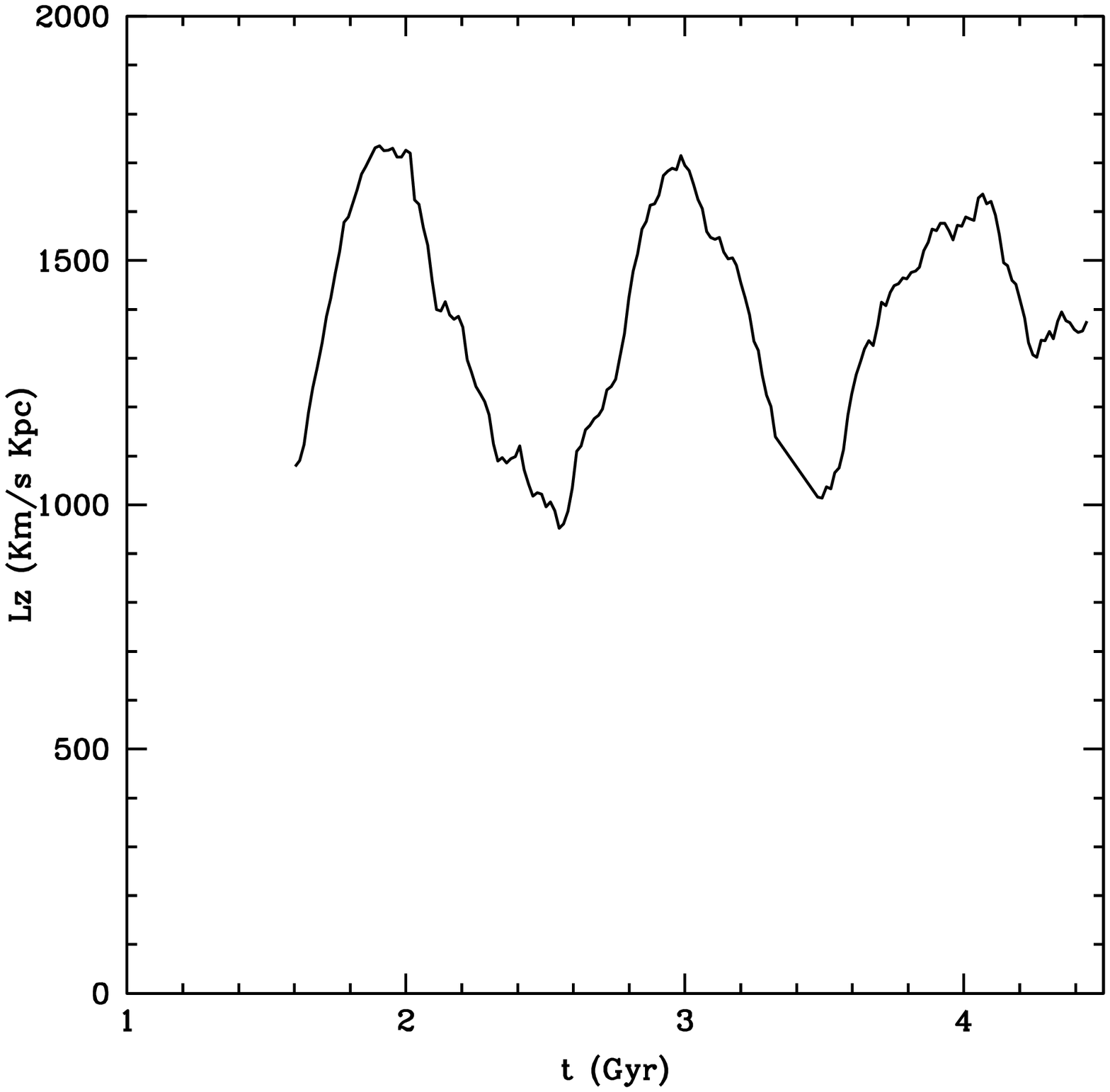}}
 \caption{ Angular momentum history of the same particle shown in figure 20. The angular momentum oscillates as the particle librates between areas of positive and negative change of angular momentum. However, there is little net change of angular momentum during the whole evolution.}
\end{figure}  
\begin{figure}
{\includegraphics[width =0.47\textwidth]{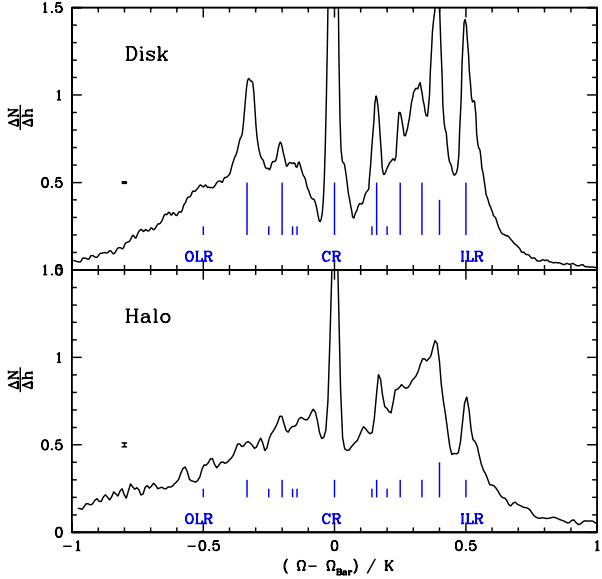}}
\caption{Bottom: Distribution of the ratio $ ( \Omega - \Omega_{B} ) /  \kappa$ for particles in the halo chosen to stay close to the disk of model 1. The lines present different resonances. The corotation and the inner Linblad resonances are clearly present in the halo. Top: the same for the disk of model 1. The errorbars in both plots are the $1\sigma$ error using poison noise. }

\section* { conclusion}
We have detected resonances in N-body simulations with evolving disks in  live halos using only trajectories of the particles. 
We find that all resonances are trapping. No gaps or scattering resonances are found. In our models, 
there are little evolution after particles get trapped.  
The corotation resonance captures particles in a ring outside the bar. Another important result is that the Inner Lindblad resonance is not localized at a given radius. 
We have also detected trapping resonances in the halo.

\end{figure}
\end{document}